# Application of the First Collision Source Method to CSNS Target Station Shielding Calculation[*]

ZHENG Ying(郑颖)[1]   ZHANG Bin(张斌)[1]   CHEN Meng-Teng(陈蒙腾)[1]

ZHANG Liang(张亮)[1]   CHEN Yi-Xue(陈义学)[1;1)]   YIN Wen(殷雯)[2]

LIANG Tian-Jiao(梁天骄)[2]

[1] School of Nuclear Science and Engineering，North China Electric Power University，Beijing 102206，China

[2] Institute of Physics, Chinese Academy of Sciences, Beijing 100190, China

**Abstract**：Ray effects are the inherent problem of discrete ordinates method. RAY3D, a functional module of ARES which is a discrete ordinates code system, employs a semi-analytic first collision source method to mitigate ray effects. This method decomposes the flux into uncollided and collided components, and then calculates them with analytical method and discrete ordinates method respectively. In this article, RAY3D is validated by the Kobayashi benchmarks and applied to the neutron beamline shielding problem of China Spallation Neutron Source (CSNS) target station. Numerical results of the Kobayashi benchmarks indicate that DONTRAN3D with RAY3D solutions agree well with the Monte Carlo solutions. The dose rate at the end of the neutron beamline is less than 10.83 μSv/h in CSNS target station neutron beamline shutter model. RAY3D can effectively mitigate ray effects and obtain relatively reasonable results.

**Key words**：CSNS, ARES, RAY3D, discrete ordinates method, first collision source method

**PACS**：28.20. Gd, 28.41.Ak

## 1. Introduction

The discrete ordinates method is the most widely used deterministic method to obtain numerical solutions of the linear Boltzmann equation and to evaluate the dose rates in nuclear devices. However, this method suffers from an inherent problem, which is called ray effects. Ray effects always cause large distortion to calculation results. In fact, the increasing of quadrature order can mitigate ray effects. However, the increase of calculation time is unacceptable. Hence, to improve the accuracy and efficiency of the discrete ordinates code system, development of a ray effects elimination module is of vital importance. In ARES code system [1], RAY3D employs the first collision source method to mitigate ray effects. RAY3D has been developed, and it is validated by the Kobayashi benchmarks and applied to CSNS target station neutron beamline shutter engineering problem.

China Spallation Neutron Source is under construction in Dongguan, and it will be the first Spallation Neutron Source facility in developing countries. Neutron beamline shutter is a critical equipment in CSNS target station. To guarantee personnel security, shielding calculation for this shielding design is very crucial. However, when using the discrete ordinates method to simulate this kind of deep penetration problem with a channel for up to 10 meters, the results always show unacceptable error because of ray effects. This paper calculates and analyzes the practical neutron beamline shutter problem using ARES

[*] Supported by Major National S&T Specific Program of Large Advanced Pressurized Water Reactor Nuclear Power Plant (2011ZX06004-007), National Natural Science Foundation of China (11275070), and the Fundamental Research Funds for the Central Universities (13QN34).

1)E-mail: yxchen1972@126.com

code system. Results obtained by DONTRAN3D with RAY3D, DONTRAN3D and TORT are presented respectively.

This paper is organized as follows. Section 2 briefly discusses the ARES code system and states first collision source theory in detail. Section 3 validates the accuracy of RAY3D by using Kobayashi benchmarks. In Section 4, we introduce the CSNS target station neutron beamline shutter model and the main calculating parameters. Section 5 analyzes the results. Section 6 concludes this article.

## 2. RAY3D code

### 2.1 Introduction of ARES code system

ARES code system, developed by School of Nuclear Science and Engineering in North China Electric Power University, is a multi-group $S_N$ particle transport code system with arbitrary order anisotropic scattering. It employs the discrete ordinates method to calculate and analyze the shielding efficiency of the nuclear devices. ARES consists of Pre, DONTRAN, RAY and Post. Pre and Post are used to preprocess and reprocess the input and output information respectively. As the deterministic solver of ARES, DONTRAN calculates the flux distribution of models accurately. Besides, RAY eliminates the ray effects by first collision source method. RAY3D is the 3-D ray effects elimination module of ARES code system. So far, ARES has been verified by authoritative benchmark problems [2] and it has been applied to many complicated nuclear facilities.

### 2.2 RAY3D and the first collision source theory

RAY3D, a three-dimensional ray effects elimination code, employs the semi-analytic first collision source theory [3] to mitigate ray effects. Ray effects, resulting from the limited discrete directions, are especially serious in systems which have void region in a highly absorbing medium. Ray effects always perform on the space oscillation of flux distribution and the enormous error of results.

First collision source method decomposes the flux into uncollided and collided components. The uncollided flux component is calculated analytically by RAY3D, and the collided flux component is calculated by DONTRAN3D.

The multigroup method is used for energy discretization in transport equation:

$$\Omega \cdot \nabla \psi^g(r,\Omega) + \Sigma_t^g(r)\psi^g(r,\Omega) = S^g(r,\Omega), \quad (1)$$

The spherical harmonics expansions are applied to the scattering source term:

$$\Omega \cdot \nabla \psi^g(r,\Omega) + \Sigma_t^g(r)\psi^g(r,\Omega) =$$
$$\sum_{g'=0}^{G}\sum_{l=0}^{N}\frac{2l+1}{4\pi}\Sigma_{sl}^{gg'}(r)[Y_{l0}^e(\Omega)\phi_{l0}^{g'}(r) + \sum_{m=1}^{l}(Y_{lm}^e(\Omega)\phi_{lm}^{g'}(r) + Y_{lm}^o(\Omega)\vartheta_{lm}^{g'}(r))]$$
$$+ q_e^g(r,\Omega),$$
$$(2)$$

Where the spherical harmonics moments of the angular flux of group g are defined by

$$\phi_{lm}^g(r) = \int_{4\pi} Y_{lm}^e(\Omega)\psi^g(r,\Omega)d\Omega, \quad m \geq 0, \quad (3)$$

$$\vartheta_{lm}^g(r) = \int_{4\pi} Y_{lm}^o(\Omega)\psi^g(r,\Omega)d\Omega, \quad m > 0, \quad (4)$$

$\phi^g$ are even order flux moments, $\vartheta^g$ are odd order flux moments.

Here
$$Y_{lm}^e(\theta,\varphi) =$$
$$(-1)^m \sqrt{(2-\delta_{m0})\frac{2l+1}{4\pi}\frac{(l-m)!}{(l+m)!}}P_{lm}(\cos\theta)\cos m\varphi,$$
$$(5)$$

$$Y_{lm}^o(\theta,\varphi) = (-1)^m \sqrt{\frac{2l+1}{2\pi}\frac{(l-m)!}{(l+m)!}}P_{lm}(\cos\theta)\sin m\varphi,$$
$$(6)$$

$P_{lm}(\cos\theta)$ are the associated Legendre polynomials.

The angular variable of the transport equation is discretized by discrete ordinates method:

$$\Omega_a \cdot \nabla \psi^g(r,\Omega_a) + \Sigma_t^g(r)\psi^g(r,\Omega_a) =$$
$$\sum_{g'=0}^{G}\sum_{l=0}^{N}\frac{2l+1}{4\pi}\Sigma_{sl}^{gg'}(r)[Y_{l0}^e(\Omega_a)\phi_{l0}^{g'}(r) + \sum_{m=1}^{l}(Y_{lm}^e(\Omega_a)\phi_{lm}^{g'}(r) + Y_{lm}^o(\Omega_a)\vartheta_{lm}^{g'}(r))] +$$
$$q_e^g(r,\Omega_a), \quad (7)$$

Where $\Omega_a$ presents a given discrete direction. $\psi^g(r,\Omega)$ are decomposed into two parts:

$$\psi^g(r,\Omega) = \psi_{(u)}^g(r,\Omega) + \psi_{(c)}^g(r,\Omega), \quad (8)$$

$\psi_{(u)}^g(r,\Omega)$ denotes the uncollided angular flux, and $\psi_{(c)}^g(r,\Omega)$ denotes the collided flux. Eq. (8) can be decomposed into the following two equations:

$$\Omega \cdot \nabla \psi_{(u)}^g(r,\Omega) + \Sigma_t^g(r)\psi_{(u)}^g(r,\Omega) = q_e^g(r,\Omega), \quad (9)$$

$$\Omega \cdot \nabla \psi_{(c)}^g(r,\Omega) + \Sigma_t^g(r)\psi_{(c)}^g(r,\Omega) =$$
$$\sum_{g'=0}^{G}\sum_{l=0}^{N}\frac{2l+1}{4\pi}\Sigma_{sl}^{gg'}(r)\sum_{m=-l}^{l}Y_{lm}(\Omega)\phi_{lm}^{g'(c)}(r) + \quad (10)$$
$$q_s^{(u)}(r,\Omega),$$

$\psi_{(u)}^g(r,\Omega)$ can be calculated analytically

$$\psi_{(u)}^g(r,\Omega) = \delta(\Omega - \Omega_{p \to r})\frac{q_0}{4\pi}\frac{e^{-\tau(r,r_p)}}{|r-r_p|^2}, \quad (11)$$

Where $r_p$ denotes the location of point source, $r$ denotes the location of targeting grid. $\tau(r,r_p)$ is the optical distance between $r$ and $r_p$. $|r-r_p|$ is the distance between $r$ and $r_p$.

The spherical harmonic moments of the uncollided angular flux become:

$$\phi_{lm}^{g(u)}(r) = Y_{lm}(\Omega_{p \to r})\frac{q_0}{4\pi}\frac{e^{-\tau(r,r_p)}}{|r-r_p|^2}, \quad (12)$$

$q_s^{(u)}(r,\Omega)$, the first collision source, can be resolved using spherical harmonics.

$$q_s^{(u)}(r,\Omega) = \sum_{g'=0}^{G}\sum_{l=0}^{N}\frac{2l+1}{4\pi}\Sigma_{sl}^{gg'}(r)\sum_{m=-l}^{l}Y_{lm}(\Omega)\phi_{lm}^{g'(u)}(r), \quad (13)$$

Firstly, RAY3D applies ray-tracing technique to obtain the numerical values of distances and optical distances between all point sources and all targeting grids respectively. Then, we use Eq. (11) to obtain the uncollided angular flux. And Eq. (12) will be used to calculate the spherical harmonic moments of the uncollided flux. The calculation of the first collision source will be accomplished by Eq. (13). The collided flux is calculated by the discrete ordinates method. The uncollided angular flux and collided flux make up total flux.

### 3. Kobayashi benchmarks

In 1996, Kobayashi proposed several 3D radiation transport benchmarks for simple geometries with void region at OECD/NEA to examine the accuracy of codes. The benchmark problems consist of three regions: the source, the void, and the shield regions, as shown in Fig. 1 and Fig. 2 [4]. Only the Problem 2 and Problem 3 are presented because the models of Problem 2 and Problem 3 are similar to CSNS target station neutron beamline shutter model.

The information of source strength and cross section are presented in table 1 [4].

Table 1. Information of source strength and cross section

| Region | $S$ (n cm$^{-3}$ s$^{-1}$) | $\Sigma_t$ (cm$^{-1}$) | case i $\Sigma_s$(cm$^{-1}$) | case ii $\Sigma_s$(cm$^{-1}$) |
|---|---|---|---|---|
| 1 | 1 | 0.1 | 0 | 0.05 |
| 2 | 0 | 10$^{-4}$ | 0 | 0.5×10$^{-4}$ |
| 3 | 0 | 0.1 | 0 | 0.05 |

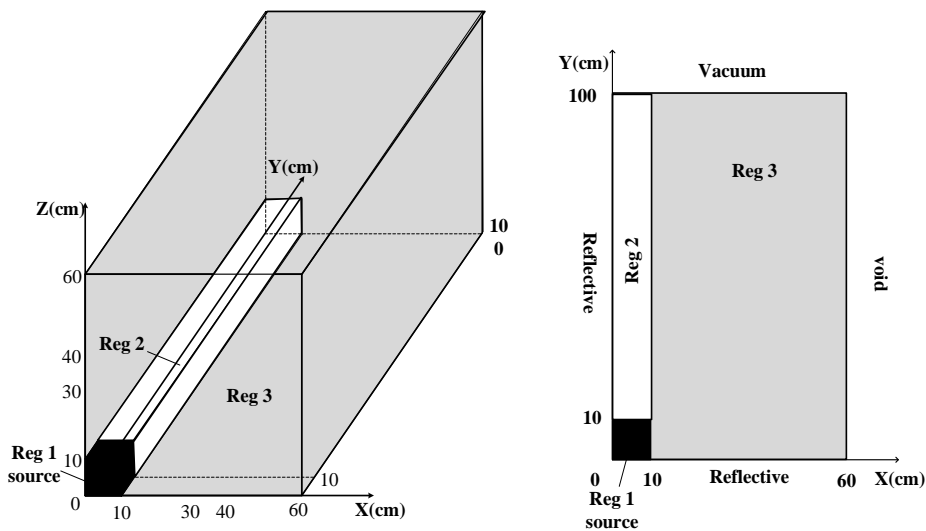

Fig. 1. Geometry of Problem 2

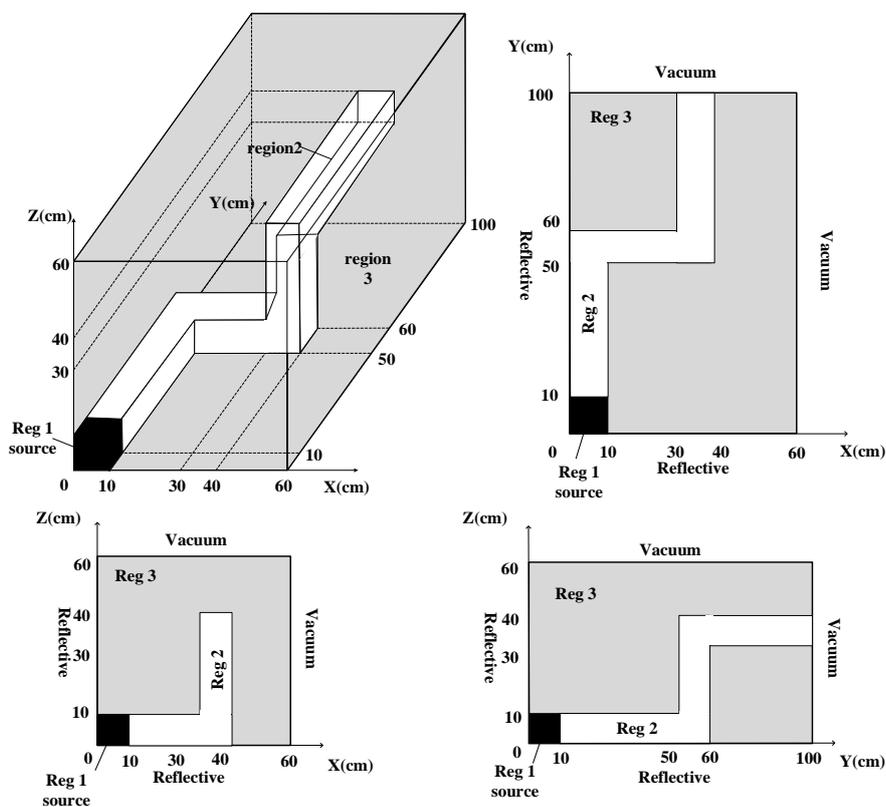

Fig.2. Geometry of Problem 3

The 3D radiation transport benchmarks were solved by Monte Carlo code MCNP4B2 in 1997 as reference, and results calculated by TORT3.2 with FNSUNCL3 in 2001 [5] are plotted for comparison. Kobayashi's benchmarks are calculated by DONTRAN and DONTRAN with RAY3D respectively. Fig. 3 and Fig. 4 plot the ratio of the calculated total neutron fluxes to that of references.

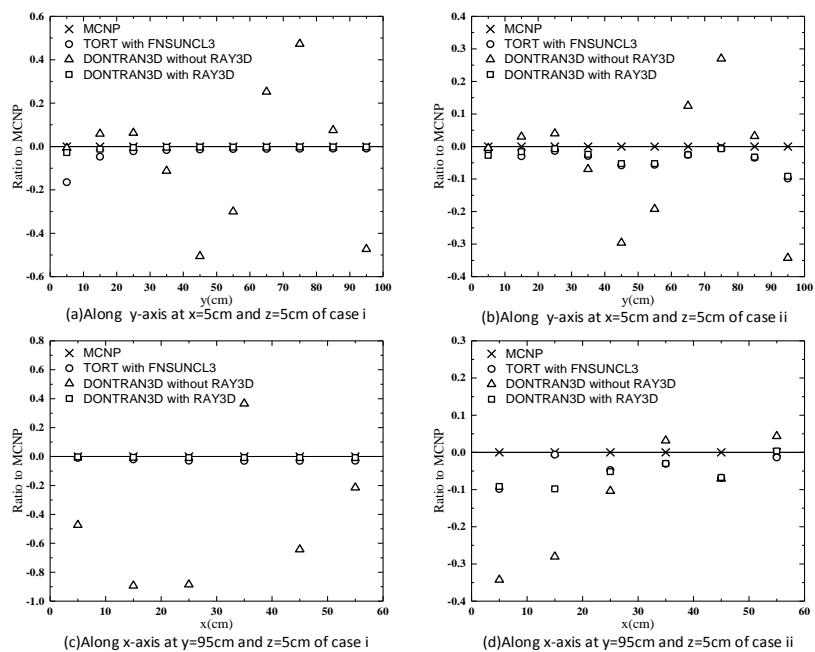

Fig. 3. Ratio of calculated total fluxes of Prob.2

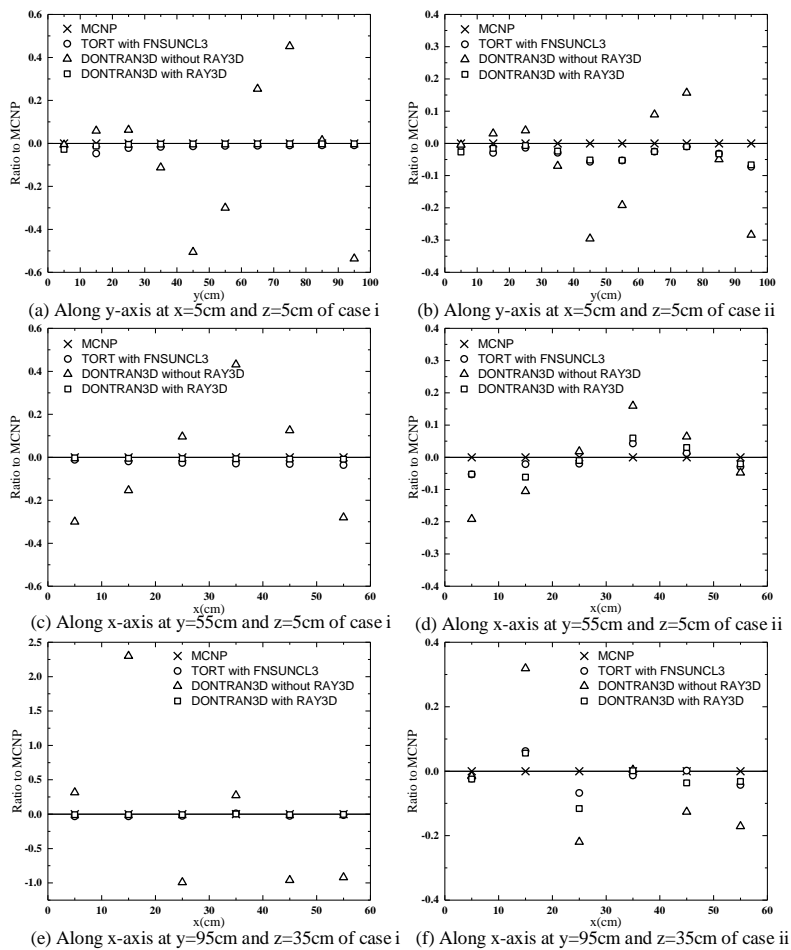

Fig. 4. Ratio of calculated total fluxes of Prob. 3

The total neutron fluxes calculated by DONTRAN3D with RAY3D show good agreement with MCNP solutions. For the cases with pure absorber, the maximum ratio is less than 4%. For the half scattering cases, the maximum ratio is less than 13%. It indicates that ARES is a reliable code system for shielding calculation.

## 4. CSNS target station neutron beamline shutter model and the calculating parameters

The CSNS target station neutron beamline shutter model [6] is shown in Fig.5. The height of model is up to 10m and the width of model is 1.15m.

The width of void region is 11cm and the void region is cut by stainless steel 316 at the height of 100cm to 450cm when the shutter is closed. Stainless steel 316 ranging from 100cm to 250cm contains a void region, as shown in Fig. 5(c). And the width of this void region is 5 cm. In Fig 5. (b), the width of stainless steel 316 ranging from 250cm to 450cm is 15cm. And the material of the rest of the model is low carbon steel. The volume of the source region is 1 cm$^3$ and it is located at the bottom-left corner in Fig 5(d). In this CSNS target station neutron beamline shutter model, boundary surfaces (X=0, Y=0 and Z=0) are treated as reflective boundary conditions. Other boundary surfaces are treated as void boundary conditions. Table 2 [6] lists the chemical compositions of stainless steel 316 and low carbon steel.

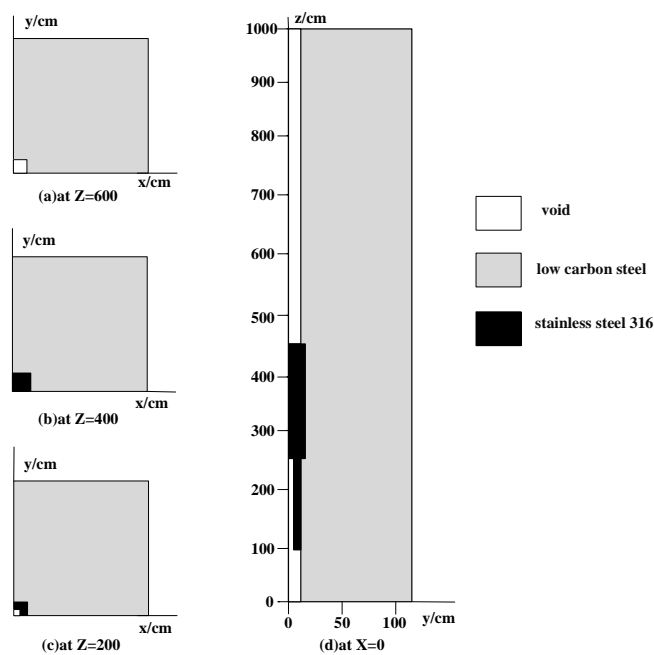

Fig. 5. The CSNS target station neutron beamline shutter model

Table 2. The material chemical compositions

| material | density | Chemical components / (wt. %) |
|---|---|---|
| stainless Steel 316 | 8.03 | iron (65.375%), chromium(17.0%), nickel(14.5%) |
| | | manganese (2.0%), silicon(1.045%), carbon(0.08%) |
| low carbon steel | 7.04 | iron(99.1%), manganese(0.45%), silicon(0.25%), carbon(0.2%) |

The model consists of 237276(39x39x156) rectangular solids. The high-energy neutrons exist in CSNS target station, so a high-energy (up to 150 MeV) multi-group library set named HEST1.0 with 253-neutron and 48-photon groups is introduced [7]. The library is developed based on ENDF/B-Ⅶ.0 using the NJOY code. We condense 253-neutron group and 48-photon group into 31-neutron group and 10-photon group respectively. $P_3$ order of the expansion in Legendre polynomials of the scattering cross-section matrix is introduced, and $S_8$ fully symmetrical quadrature sets which approximate the flux angular discretization is employed.

## 5. Results and analysis

To illustrate that the ray effects are the common problems of discrete ordinates codes and they are unacceptable in CSNS target station neutron beamline shutter problem, we present the distribution of dose rate at the height of 0 to 1000cm (x=0cm, y=0cm) using TORT and DONTRAN3D respectively. Then we calculate the same model using DONTRAN3D with RAY3D. The results of dose rate distribution at the height of 0 to 1000cm (x=0, y=0) are shown in Fig. 6.

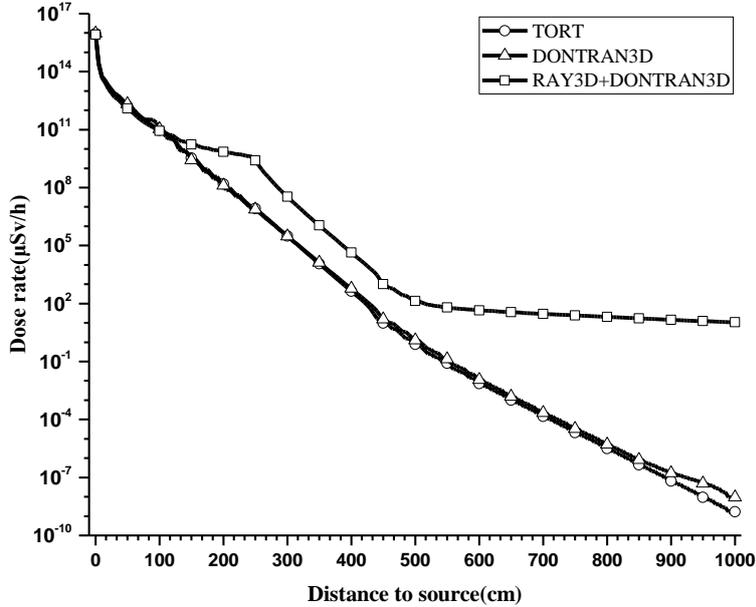

Fig. 6. The total dose axial distribution using TORT and DONTRAN

As shown in Fig. 6, from z=0cm to z=450cm, results obtained by TORT are nearly the same as results calculated by DONTRAN3D. At the height of 450cm to 1000cm, which is a region without the strong shielding action of stainless steel 316, results of TORT and DONTRAN3D decline rapidly along z-axis. Results of DONTRAN3D appear to be an order of magnitude greater than that of TORT at z=1000cm. The difference is caused by the different difference scheme. In DONTRAN3D, DZ (diamond set to zero difference scheme) is employed. While WD (weighted difference scheme) is employed in TORT. The unreasonable decline indicates that it is infeasible to solve CSNS target station neutron beamline shutter problem just using discrete ordinates code. In this kind of problem with void region and small source region, ray effects are especially serious. As can be seen in Fig. 6, results of DONTRAN3D with RAY3D are obviously different. At the height of 250cm to 450 cm, dose rate decline rapidly along z-axis because of the shielding of stainless steel 316. The most obvious difference between results of DONTRAN3D and DONTRAN3D with RAY3D appears at the height of 450cm to 1000cm. In this void region, the result of DONTRAN3D with RAY3D decline much smoother along z-axis. So, with the ray effects elimination effect of RAY3D, the results are more reasonable. And the dose rate at z=1000cm is 10.83 μSv/h.

To intuitively show the comparison of results calculated by DONTRAN3D and DONTRAN3D with RAY3D, the distribution of fluxes in an arbitrary group is chosen to be presented. Fig. 7 provides contour plots of the scalar flux of energy group 1 along the y-z plane at x=0. And Fig. 8 provides contour plots of the total dose rate along the y-z plane at x=0 using DONTRAN3D and DONTRAN3D with RAY3D.

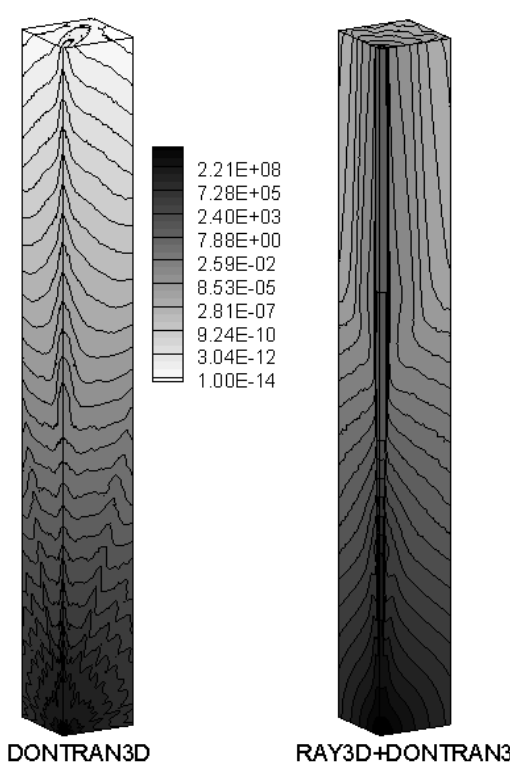

Fig. 7. Scalar flux distribution of energy group 1 along the y-z plane at x=0

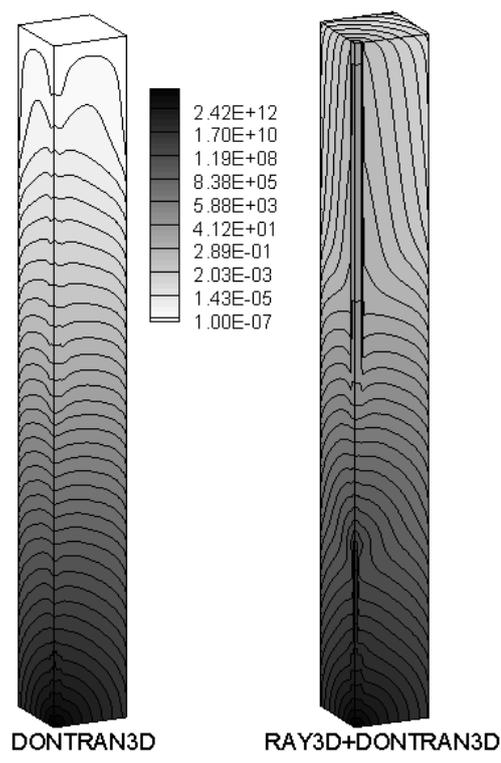

Fig. 8. Total rate dose distribution along the y-z plane at x=0

## 6. Conclusions

RAY3D, which employs the first collision source method, can effectively mitigate ray effects. And ARES code system can effectively solve CSNS target station neutron beamline shutter problem. The dose rate at the end of the neutron beam line is 10.83 μSv/h. Thus ARES provides an accurate tool for researchers to optimize a best engineering design in order to maintain a safe working environment and to minimize the cost of shielding materials.

There is still further work to do to make our code system perform better, like using the angle-integrated mesh-cell balance equation to modify the first collision source method [8] and developing a library with higher upper limit of energy.

*The authors gratefully thank all members of CSNS Neutronics Group for their valuable work.*